\input{aipcheck}

\documentclass[
    ,final            
  ]
  {aipproc}

\layoutstyle{6x9}

\begin{document}

\title{Particle phenomenology and Maldacena}



\author{Paul H. Frampton}{
  address={Department of Physics and Astronomy,
  University of North Carolina, Chapel Hill, NC 27599, USA.} }

\begin{abstract}
A brief review is offered of employing Maldacena's AdS/CFT
correspondence in attempting to identify a model
which extends to higher energy the standard model of particle phenomenology.
\end{abstract}

\maketitle


\section{Introduction}

It is an honor to speak at a meeting in his home
city commemorating
the tenth anniversary of Maldacena's 
paper\cite{Maldacena}
on the AdS/CFT correspondence
which has acquired five thousand citations
in its first ten years.
Its popularity stems from the diversity of
its applications. As merely one example, it attracted me
back to string theory in 1998 after years away by a 
derivative conjecture\cite{Frampton},
admittedly going beyond what Maldacena 
stated, that a non-supersymmetric finite $SU(N)$ gauge
theory can be conformally invariant at high energies, 
a conjecture which has the disadvantage
of being technically difficult to study.
A no-go theorem\cite{Dymarsky}
exists but uses
questionable assumptions.

\section{AdS/CFT and standard model}

The physical idea is to consider the standard
model from a perspective of an energy of a few TeV
or more where quark and lepton masses, the QCD and weak scales,
become negligible and the theory is classically scale invariant.
Quantum mechanically it is not so because of beta functions
and anomalous dimensions which generically
do no vanish. The idea is to enrich the theory in such
a way that it becomes also quantum mechanically conformally invariant at
high energies.

\bigskip

A recent comprehensive review\cite{FKreview} contains a
historical section which indicates the special role
played by the paper\cite{Maldacena} 
in phenomenology beyond the standard model and it seems 
appropriate to reproduce it in its entirety here.

\bigskip

\section{Impact of Maldacena's paper on particle phenomenology}

\bigskip

Particle phenomenology is in an especially exciting time, mainly
because of the anticipated data in a new energy regime expected
from the Large Hadron Collider (LHC), to be commissioned at the
CERN Laboratory in 2007. This new data is long overdue. The 
Superconducting Supercollider (SSC) could have provided such data
long ago were it not for its political demise in 1993.

Except for the remarkable experimental data concerning neutrino
masses and mixings which has been obtained since 1998, 
particle physics has been data starved for the last thirty years.
The standard model invented in the '60s and '70s has been confirmed and
reconfirmed. Consequently, theory has ventured into speculative
areas such as string theory, extra dimensions and supersymmetry.
While these ideas are of great interest and theoretically consistent
there is no {\it direct} evidence from experiment for them.
Here we describe a more recent, post 1998, direction known as
conformality. First, to set the stage, we shall discuss why
the conformality approach which is, in our opinion, competitive
with the other three approaches, remained unstudied
for the twenty years up to 1998.

A principal motivation underlying model building, beyond the
standard model, over the last thirty years has been the
{\it hierarchy problem} which is a special case of {\it naturalness}.
This idea stems from Wilson\cite{Wilson1971} in the late '60s.
The definition of naturalness is that a theory should not contain
any unexplained very large (or very small in the inverse)
dimensionless numbers. The adjustment needed to achieve such
naturalness violating numbers is called {\it fine tuning}.
The naturalness situation can be especially acute in gauge
field theories because even after fine tuning at tree level,
{\it i.e.}, the classical lagrangian, the fine tuning may need to
be repeated an infinite number of times order by order
in the loop expansion during the renormalization process.
While such a theory can be internally consistent it
violates naturalness. Thus naturalness is not only an aesthetic
criterion but one which the vast majority of the community
feel must be imposed on any acceptable extension of the
standard model; ironically, one exception is Wilson himself
\cite{Wilson2004}.

When the standard model of Glashow\cite{Glashow1961} was rendered
renormalizable by appending the Higgs mechanism\cite{Weinberg1967,Salam1968}
it was soon realized that it fell into trouble with naturalness,
specifically through the hierarchy problem. In particular, the
scalar propogator has quadratically divergent radiative corrections
whch imply that a bare Higgs mass $m_H^2$ will be corrected
by an amount $\Lambda^2/m_H^2$ where $\Lambda$ is the cut off
scale corresponding to new physics. Unlike logarithmic
divergences, which can be absorbed in the usual renormalization
process, the quadratic divergences create an unacceptable
fine tuning: for example, if the cut off is at the conventional
grand unification scale $\Lambda \sim 10^{16}$ GeV and $m_H \sim 100$
GeV, we are confronted with a preposterous degree of fine tuning
to one part in $10^{28}$.

As already noted, this hierarchy problem was stated most forcefully
by Wilson who said, in private discussions, that scalar fields
are forbidden in gauge field theories. Between the late '60s and 1974,
it was widely recognized that the scalar fields of the standard model
created this serious hierarchy problem but no one knew what to do
about it.

The next big progress to the hierarchy problem came in 1974 with
the invention\cite{Wess1974} of supersymmetry. This led to the Minimally Supersymmetric
Standard Model (MSSM) which elegantly answered Wilson's objection
since quadratic divergences are cancelled between bosons and
fermions, with only logarithmic divergences surviving. Further it was
proved \cite{Haag1975,Haag19752} that the MSSM and straightforward
generalizations were the unique way to proceed.
Not surprisingly, the MSSM immediately became
overwhelmingly popular. It has been estimated \cite{Woit2006}
that there are 35,000 papers existing on supersymmetry, more
than an average of one thousand papers per year since its
inception. This approach continued to seem "unique" until 1998. Since
the MSSM has over one hundred free parameters, many
possiblities needed to be investigated and exclusion plots
constructed. During this period, two properties beyond
naturalness rendered the MSSM even more appealing: an
improvement in unification properties and a candidate
for cosmological dark matter.

Before jumping to 1998, it is necessary to mention an unconnected
deveopment in 1983 which was the study of Yang-Mills theory
with extended ${\cal N}=4$ supersymmetry (the MSSM has
${\cal N}=1$ supersymmetry). This remarkable theory, though
phenomenologically quite unrealistic as it allows no
chiral fermions and all matter fields are in adjoint representions,
is finite
\cite{Mandelstam1983,Mandelstam19832,Mandelstam19833}
to all orders of perturbation theory
and conformally invariant. Between 1983 and 1997, the relationship
between the ${\cal N}=4$ gauge theory and either string theory,
also believed to be finite, or the standard model remained unclear.

The perspective changed in 1997-98
initially through the insight of Maldacena\cite{Maldacena}
who showed a {\it duality} between ${\cal N} = 4$ gauge theory
and the superstring in ten spacetime dimensions. Further
the ${\cal N} = 4$ supersymmetry can be broken by orbifolding
down to ${\cal N} = 0$ models with no supersymmetry at all.
It was conjectured \cite{Frampton}
by one of the authors in 1998
that such nonsupersymmetric orbifolded models can be finite
and conformally invariant, hence the name conformality.

Conformality models have been investigated far less completely
than supersymmetric ones but it is already
clear that supersymmetry is ``not as unique'' as previously
believed. No-go theorems can have not only explicit
assumptions which need to be violated to avoid the theorem
but unconcious implicit assumptions which require further
progress even to appreciate: in 1975 the implicit assumption
was that the gauge group is simple, or if semi-simple
may be regarded as a product of theories each with a simple
gauge group. Naturalness, by cancellation of quadratic
divergences, accurate unification and a dark matter
candidate exist in conformality.

It becomes therefore a concern that the design of the
LHC has been influenced by the requirement of testing
the MSSM. The LHC merits
an investment of theoretical work to check if the
LHC is adequately designed to test conformality which now
seems equally as likely as supersymmetry, although we
fully expect the detectors
ATLAS and CMS to be sufficiently
all purpose to capture any
physics beyond the standard model at the TeV scale.

\bigskip

\section{Quiver gauge theories}

\bigskip

Quiver gauge theories possess a gauge group which is generically
a product of $U(N_i)$ factors with matter fields in bifundamental
representations. They have been studied in the physics literature
since the 1980s where they were used in composite model building.
They have attracted much renewed attention because of their
natural appearance in the duality between superstrings and gauge theories.

The best known such duality gives rise to a highly supersymmetric
(${\cal N}=4$) gauge theory with a single $SU(N)$ gauge group with matter
in adjoint representations. In this case one can drop with impunity
the $U(1)$ of $U(N)$ because the matter fields are uncharged under it.
In the quiver theories with less supersymmetry (${\cal N} \le 2$)
it is usually necessary to keep such $U(1)$s.  

Quiver gauge theories are taylor made for particle physics model building. While an $SU(N)$
gauge theory is typically anomalous in  for arbitrary choice of fermions,  
choosing the fermions to lie in a quiver insures anomaly cancelation. 
Furthermore the fermions in a quiver arrange themselves
in bifundamental representations  of the product gauge group. This nicely coinsides with the fact
that all known fundamental fermions are in bifundamental, fundamental, or 
singlet representations of the gauge group. The study of quiver gauge  
theories goes back to the earliest days of gauge theories and the standard
model. Other notable early examples are the Pati-Salam model  
and the trinification model. A vast literature exists on 
this subject, but we will concentrate on post  
$AdS/CFT$ conjecture quiver gauge theory work
\cite{FKreview}
Starting from $AdS_5\times S^5$ we only have an $SU(N)$ ${\cal N}=4 $ supersymmetric
gauge theory. In order to break SUSY and generate a quiver gauge theory 
there are several options open to us. 
Orbifolds \cite{FKreview},
conifolds \cite{FKreview}
and orientifolds \cite{FKreview}
have all played a part in building quiver gauge theories.

\bigskip

It is important to note that although the duality with superstrings
is a significant guide to such model building, and it is desirable
to have a string dual to give more confidence in consistency, we
shall focus on the gauge theory description in the approach to
particle phenomenology, as there are perfectly good quiver gauge
theories that have yet to be derived from string duality.

\section{Orbifolding}

\bigskip

The simplest superstring - gauge duality arises from the compactifiation
of a Type IIB superstring on the cleverly chosen manifold

\bigskip

\[ ~~~~~~~ AdS_5 ~~ \times ~~ S^5  \]

\bigskip

\noindent which yields an ${\cal N} = 4$ supersymmetry which is an especially
interesting gauge theory which has been intensively studied and possesses
remarkable properties of finiteness and conformal invariance for all
values of $N$ in its $SU(N)$ gauge group. By ''conformality", we shall
mean conformal invariance at high energy, also for finite $N$.

For phenomenological purposes, ${\cal N} = 4$ is too much supersymmetry.
Fortunately it is possible to breaking supersymmetries and hence
approach more nearly the real world, with less or no supersymmetry in a
conformality theory.

By factoring out a discrete (either abelian or nonabelian) group and composing an orbifold:

\bigskip

\[ ~~~~~~S^5 / \Gamma ~~~~~~~\]

\bigskip

\noindent one may break ${\cal N} = 4$ supersymmetry to
${\cal N} = 2, ~~~~1,$ or $~~~0$. Of special interest is the ${\cal N} = 0$ case.

We may take an abelian $\Gamma = Z_p$ (non-abelian cases will also be considered
in this review) which identifies $p$ points in a complex three dimensional space ${\cal C}_3$.

The rules for breaking the ${\cal N} = 4$ supersymmetry are:

\bigskip

\noindent If $\Gamma$ can be embedded in an $SU(2)$ of the original $SU(4)$ R-symmetry, then
\[
~~ \Gamma \subset SU(2)~~~~\Rightarrow~~{\cal N} = 2.
\]

\noindent If $\Gamma$ can be embedded in an $SU(3)$ but not an $SU(2)$ of the original $SU(4)$ R-symmetry, then
\[
~~ \Gamma \subset SU(3)~~~~\Rightarrow~~{\cal N} = 1.
\]

\noindent If $\Gamma$ can be embedded in the $SU(4)$ but not an $SU(3)$ of the original $SU(4)$ R-symmetry, then
\[
~~ \Gamma \subset SU(4)~~~~\Rightarrow~~{\cal N} = 0.
\]

\bigskip

\noindent In fact to specify the embedding of $\Gamma = Z_p$ we need to identify three integers $(a_1, a_2, a_3)$:

\[
~ {\cal C}_3 :~~(X_1, X_2, X_3)~~\stackrel{Z_p}{\rightarrow}~(\alpha^{a_1} X_1, \alpha^{a_2} X_2, \alpha^{a_3}X_3)
\]

\noindent with

\[
~ \alpha = exp \left( \frac{2 \pi i}{p} \right)
\]

\noindent The $Z_p$ discrete group identifies $p$ points in ${\cal C}_3$.
The N converging D3-branes meet on all $p$ copies, giving a gauge group:
$U(N) \times U(N) \times ......\times U(N)$, $p$ times.
The matter (spin-1/2 and spin-0)
which survives is invariant
under a product of a
gauge transformation and a $Z_p$ transformation.

There is a convenient diagramatic way to find the result from
a ''quiver."   One draws $p$ points and arrows for $a_1, a_2, a_3$.

\bigskip

\noindent For a general case, the scalar representation contains the bifundamental
scalars

\bigskip

\[\sum_{k=1}^{3}\sum_{i = 1}^{p} (N_1, \bar{N}_{i \pm a_k})\]

\bigskip

For fermions, one must first construct the {\bf 4} of R-parity $SU(4)$, isomorphic to
the isometry $SO(6)$ of the $S^5$.
From the $a_k = (a_1, a_2, a_3)$ one constructs the 4-spinor $A_{\mu} = (A_1, A_2, A_3, A_4)$ :

\bigskip

\[ A_1 = \frac{1}{2} (a_1 + a_2 +a_3) \]

\[ A_2 = \frac{1}{2} (a_1 - a_2 -a_3) \]

\[ A_3 = \frac{1}{2} (- a_1 + a_2 - a_3) \]

\[ A_4 = \frac{1}{2} (- a_1 - a_2 +a_3) \]

\noindent These transform
as $exp \left( \frac{2 \pi i}{p} A_{\mu} \right)$ and the
invariants may again be derived by a different quiver diagram. 

\bigskip

\noindent Note that these lines are oriented, as is necessary to accommodate chiral
fermions. Specifying the four $A_{\mu}$ is equivalent (there is a constraint that the
four add to zero, mod $p$) to fixing the three $a_k$ and group theoretically is more fundamental.

\bigskip

\noindent  In general,  the fermion representation contains the
bifundamentals

\bigskip

\[ \sum_{\mu = 1}^{4} \sum_{i = 1}^{p} ( N_i, \bar{N}_{i + A_{\mu}}) \]

\noindent When one of the $A_{\mu}$s is zero, it signifies a degenerate case of a bifundamental
comprised of adjoint and singlet representations of one $U(N)$.

\bigskip

\section{Conformality phenomenology}

\bigskip

In attempting to go beyond the standard model, one outstanding issue is
the hierarchy between GUT scale and weak scale
which is 14 orders of magnitude. Why do these two very different
scales exist?
Also, how is this hierarchy of scales stabilized
under quantum corrections?
Supersymmetry answers the second question but not the first.

The idea is to approach hierarchy problem by Conformality at a TeV Scale.
We will show how this is possible including explicit examples containing standard model states.

In some sense conformality provides more rigid constraints than supersymmetry.
It can predict additional states at TeV scale, while there can be far fewer initial
parameters in conformality models than in SUSY models.
Conformality also provides a new approach to gauge coupling unification.
It confronts naturalness and provides cancellation of quadratic divergences.
The requirements of anomaly cancellationsi can lead
to conformality of U(1) couplings.

There is a viable dark matter candidate, and proton decay
can be consistent with experiment.

\bigskip

What is the physical intuition and picture underlying conformality?
Consider going to an energy scale higher than the weak scale,
for example at the TeV scale. Quark and lepton masses,
QCD and weak scales small compared to TeV scale. They may
be approximated by zero. The theory is then classically
conformally invariant though not at the quantum level because
the standard model has non-vanishing renormalization group
beta functions and anomalous dimensions. So this suggests that we
add degrees of freedom to yield  a gauge field theory
with conformal invariance.
There will be 't Hooft's naturalness since the zero mass limit
increases symmetry to conformal symmetry.

\bigskip

We have no full understanding of how four-dimensional
conformal symmetry can be broken spontaneously so
breaking softly by relevant operators is a first step.
The theory is assumed to be given by the action:

\bigskip

\begin{equation}
S = S_0 + \int d^4x \alpha_i O_i
\end{equation}

\bigskip

\noindent where $S_0$ is the action for the conformal theory and the $O_i$ are
operators with dimension below four ({\it i.e.} relevant)
which break conformal invariance softly.

\bigskip

\noindent The mass parameters $\alpha_i$ have mass dimension $4-\Delta_i$ where
$\Delta_i$ is the dimension of $O_i$ at the
conformal point.

\bigskip

\noindent Let $M$ be the scale set by the parameters $\alpha_i$ and
hence the scale at which conformal invariance is broken. Then for $E >> M$ the couplings
will not run while they start running for $E < M$.
To solve the hierarchy problem we assume $M$ is near the TeV scale.

\bigskip

\section{Experimental evidence for conformality}

Consider embedding the standard model gauge group according to:

\[ SU(3) \times SU(2) \times U(1) \subset \bigotimes_i U(Nd_i) \]

Each gauge group of the SM can lie entirely in a $SU(Nd_i)$
or in a diagonal subgroup of a number thereof.

\bigskip

\noindent Only bifundamentals (including adjoints) are possible.
This implies no $(8,2), (3,3)$, etc. A conformality restriction which is new and
satisfied in Nature! The fact that the standard model has matter
fields all of which can be accommodated as bifundamentals
is expermental evidence for conformality.

\bigskip

\noindent No $U(1)$ factor can be conformal in perturbation
theory and so hypercharge is quantized
through its incorporation in a non-abelian gauge group.
This is the ``conformality'' equivalent to the GUT charge quantization
condition in {\it e.g.} $SU(5)$. It can explain the neutrality of the hydrogen atom.
While these are postdictions, the predictions of the theory are new particles,
perhaps at a low mass scale,  filling out
bifundamental representations of the gauge group that restore conformal
invariance. The next section will begin our study of known quiver
gauge theories from orbifolded $AdS^5\times S^5$.

\bigskip

\section{Tabulation of the simplest abelian quivers}

\bigskip

We consider the compactification of the type-IIB superstring
on the orbifold $AdS_5 \times S^5/\Gamma$
where $\Gamma$ is an abelian group $\Gamma = Z_p$
of order $p$ with elements ${\rm exp} \left( 2 \pi i A/p \right)$,
$0 \le A \le (p-1)$.

The resultant quiver gauge theory has ${\cal N}$
residual supersymmetries with ${\cal N} = 2,1,0$ depending
on the details of the embedding of $\Gamma$
in the $SU(4)$ group which is the isotropy
of the $S^5$. This embedding is specified
by the four integers $A_m, 1 \le m \le 4$ with

\begin{equation}
\Sigma_m A_m = 0~ {\rm mod } ~p
\nonumber
\label{SU4}
\end{equation}

\noindent which characterize
the transformation of the components of the defining
representation of $SU(4)$. We are here interested in the non-supersymmetric
case ${\cal N} = 0$ which occurs if and only if
all four $A_m$ are non-vanishing.

\bigskip

Table I.  All abelian quiver
theories with ${\cal N}=0$ from $Z_2$ to $Z_5$.

\bigskip

\begin{tabular}{|||c||c||c|c||c|c|c||c|c|||}
\hline
\hline
& p & $A_m$ & $a_i$ & scal & scal & chir &  \\
&&&& bfds & adjs & frms & SM  \\
\hline
\hline
1 & 2 & (1111) & (000) & 0 & 6 & No  &  No \\
\hline
\hline
2 & 3 & (1122) & (001) & 2 & 4 & No  &  No \\
\hline
\hline
3 & 4 & (2222) & (000) & 0 & 6 & No  &  No \\
4 & 4 & (1133) & (002) & 2 & 4 & No  &  No \\
5 & 4 & (1223) & (011) & 4 & 2 & No  &  No \\
6 & 4 & (1111) & (222) & 6 & 0 & Yes  &  No \\
\hline
\hline
7 & 5 & (1144) & (002) & 2 & 4 & No  &  No \\
8 & 5 & (2233) & (001) & 2 & 4 & No  &  No \\
9 & 5 & (1234) & (012) & 4 & 2 & No  &  No \\
10 & 5 & (1112) & (222) & 6 & 0 & Yes  &  No \\
11 & 5 & (2224) & (111) & 6 & 0 & Yes  &  No \\
\hline
\hline
\end{tabular}

\bigskip

\section{Chiral fermions}

The gauge group is $U(N)^p$. The fermions
are all in the bifundamental representations
\begin{equation}
\Sigma_{m=1}^{m=4}\Sigma_{j=1}^{j=p} (N_j, \bar{N}_{j + A_m})
\label{fermions}
\end{equation}
which are manifestly non-supersymmetric because no
fermions are in adjoint representations
of the gauge group.
Scalars appear in representations
\begin{equation}
\Sigma_{i=1}^{i=3}\Sigma_{j=1}^{i=p} (N_j, \bar{N}_{j \pm a_i})
\label{scalars}
\end{equation}
in which the six integers $(a_i, -a_i)$ characterize the
transformation of the
antisymmetric second-rank tensor representation
of $SU(4)$. The $a_i$
are given by $a_1 = (A_2+A_3), a_2= (A_3+A_1)$, and $a_3= (A_1+A_2)$.

It is possible for one or more of the $a_i$ to vanish
in which case the corresponding scalar representation
in the summation in Eq.(\ref{scalars}) is to be interpreted as an
adjoint
representation of one particular $U(N)_j$.
One may therefore
have zero, two, four or all six of the scalar
representations, in Eq.(\ref{scalars}), in such adjoints.
One purpose of the present article is to
investigate how the renormalization properties
and occurrence of quadratic divergences
depend on the distribution
of scalars into bifundamental
and adjoint representations.

Note that there is one model with all scalars in adjoints for each even
value of $p$. For general even $p$
the embedding is
$A_m=(\frac{p}{2},\frac{p}{2},\frac{p}{2},\frac{p}{2})$. This series
by itself forms the complete list of ${\cal N}=0$ abelian quivers with
all scalars in adjoints.

To be of more phenomenolgical interest the model should
contain chiral fermions. This requires that the embedding
be complex: $A_m \not\equiv -A_m$ (mod p). It will now be shown
that for
the presence of chiral fermions all scalars must be in bifundamentals.

The proof of this assertion follows by assuming the contrary,
that there is at least one adjoint arising from, say, $a_1=0$.
Therefore
$A_3=-A_2$ (mod p). But then it follows from Eq.(\ref{SU4})
that $A_1=-A_4$ (mod p). The fundamental representation of $SU(4)$
is thus real and fermions are non-chiral.

The converse also holds: If all $a_i \neq 0$ then there are chiral
fermions.
This follows since by assumption
$A_1 \neq -A_2$, $A_1 \neq -A_3$, $A_1 \neq -A_4$. Therefore
reality of the fundamental representation would require
$A_1 \equiv -A_1$ hence, since $A_1 \neq 0$, $p$ is even
and $A_1 \equiv \frac{p}{2}$; but then the other $A_m$
cannot combine to give only vector-like fermions.

It follows that:

\bigskip

\noindent \underline{{\it In an ${\cal N}=0$ quiver gauge theory, chiral fermions
are possible}}

\noindent \underline{{\it if and only if all scalars are in bifundamental
representaions.}}

\bigskip

\section{other developments}

\bigskip

The orbifold model buiding has been extended to non-abelian
finite groups including the analysis of ebery such froup
of order $g \leq 31$. This can give rise to more general
unifying gauge groups like $SU(4) \times SU(2) \times SU(2)$
and to interesting such chiral models.

\bigskip

Grand unification models with TeV scale unification haev attracted
attention because the unification of the couplings occurs in a
novel fashion associated with the group embeddings
of $SU(3), SU(2)$ and $U(1)$. Such unification is precisely accurate,
as much so or more than supersymmetric grand unification.

\bigskip

Quadratic diverences in the scalar two-point function are
canceled due to a generalization of supersymmetry,
named misaligned supersymmetry whose explicit realization is 
a challenging open question.

\bigskip

There is an attractive dark matter candidate called the LCP or Lightest Conformality
Particle.

\bigskip

For more details about these further developments we refer the reader
to the review article listed\cite{FKreview}.

\bigskip

\section{Congratulations}

\bigskip

When the paper \cite{Maldacena} first appeared in November 1997,
having not recently worked on string theory,
I remained unaware of it until July
1998 when, visiting CERN, almost every theory seminar was about AdS/CFT. 
There had recently been a string conference in Santa Barbara
where almost all talks were on AdS/CFT. Participants there even danced 
to an AdS/CFT song\cite{UCSB}!

\bigskip

I have talked about physics a number of times with Maldacena
who shares, with {\it e.g.} Nambu, exceptional modesty. 
Administration may nurture
his single-processor thinking, a bit reminiscent of Einstein and 
general relativity?
It has been stimulating to write papers
about AdS/CFT and, ignoring the admonition that
each self-citation counts (-5), here is my
list \cite{selfcite}.

\bigskip

\noindent CONGRATULATIONS TO AdS/CFT ON ITS TENTH ANNIVERSARY

\bigskip

\section*{Acknowledgements}

It is a pleasure to thank the organizers of this festschrift
for a memorable meeting in Buenos Aires.
This work was supported in part by the U.S. Department of Energy
under grant No. DE-FG02-06ER41418.

\end{document}